\documentstyle[aaspp4,12pt,flushrt]{article}

\def\ifm#1{\relax\ifmmode#1\else$\mathsurround=0pt #1$\fi}

\def\degrees{\hbox{${}^\circ$\hskip-3pt .}}

\begin{document}

\title{\Large\bf A Characteristic Scale on the Cosmic Microwave Sky}

\author{\large\bf Elena Pierpaoli\altaffilmark{1,2},
Douglas Scott\altaffilmark{1}
\& Martin White\altaffilmark{3}}

{\small \noindent $^1$Department of Physics and Astronomy, University of
British Columbia, B.C.\ V6T 1Z1, Canada}\\
{\small \noindent $^2$Canadian Institute for Theoretical Astrophysics,Toronto,
ON\ M5S 3H8, Canada}\\
{\small \noindent $^3$Harvard-Smithsonian Center for Astrophysics,
Cambridge, MA 02138, U.S.A.}

The 1965 discovery (1) of the Cosmic Microwave Background (CMB) was key
evidence supporting the hot Big Bang model for the evolution of the Universe.
The tiny temperature variations discovered in 1992 (2) -- of just the right
size for gravity to have grown the observed large-scale structures over the
age of the Universe -- established gravitational instability as the mechanism
of structure formation.
Those first measurements of CMB anisotropy on tens of degree scales have been
followed by many experiments concentrating on smaller angular scales.
Even 5 years ago (3) there were indications for enhanced temperature
variations on half-degree scales.
By combining results from all current experiments it is now clear that this
`excess power' decreases again below half a degree -- in other words there is
a distinctive scale imprinted upon the microwave sky.
The existence of such a feature at roughly $0\degrees5$ has profound
implications for the origin of structure in the Universe and the global
curvature of space.

It is conventional to expand the CMB sky into a set of orthogonal basis
functions labeled by `multipole number' $\ell$.  Functions with higher
$\ell$ probe smaller angular scales.
We then consider the squares of the expansion coefficient amplitudes as a
function of $\ell$, or inverse angle, and this is referred to as the
`anisotropy power spectrum' (4).
This power spectrum is easy to compute theoretically, and in popular models
contains essentially all of the cosmological information in the CMB.

What remains is to obtain this power spectrum experimentally.
Each experiment is sensitive to a range of angular scales, and its sensitivity
as a function of $\ell$ is encoded in its `window function'.
Several experiments can now divide their
$\ell$ range into overlapping window functions and thus obtain information on
the shape of the power spectrum.  Each experiment thus quotes results for one
or more `band-powers', which is the amplitude of the anisotropies
integrated over the window function (5).
Individual experiments until now have had limited angular range, so each has
provided only a small piece of the puzzle.  However a number of different
CMB experiments can be combined together to provide an essentially
model-independent estimate of the power spectrum.
This estimate, provided it is carefully calculated, can then be used
to constrain models.

We used a maximum likelihood technique to combine the band-powers into a
binned power spectrum encapsulating the knowledge gained from the different
observations.
We have included all the experimental results of which we are currently aware.
Specifically those collected in Ref.~(6), together with the more
recent results of the QMAP (7), MAT (8), Viper (9) and BOOM97 (10)
experiments; as summarized in the {\sc Radpack} package (11) with some minor
corrections.

\begin{figure}
\begin{center}
\leavevmode
\centerline{\epsfysize=10cm \epsfbox{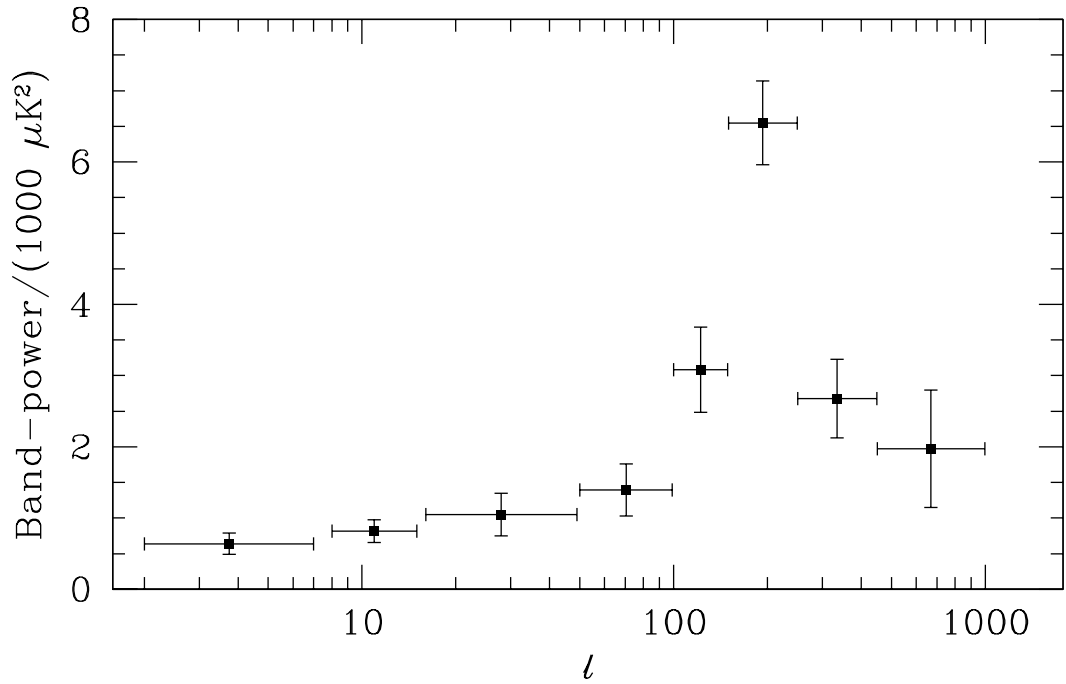}}
\end{center}
\caption{The power spectrum of Cosmic Microwave Background anisotropies.
This is a plot of temperature variations versus multipole, which is the
equivalent of an inverse angle.  The plot is a binned spectrum from all
the currently available data.  There is clearly a peak which is localized
in angle.}
\label{fig:bandpower}
\end{figure}

For definiteness we have divided the range $\ell=2$--$1000$ into 8 bins
(spaced at roughly equal logarithmic intervals, with slight adjustment to
allow for regions where data are scarcer).
As the experimental situation improves, particularly at higher $\ell$, we
expect that emphasis will shift to plots linear in $\ell$ and having a wider
range -- however, for now the situation is adequately summarized in a log plot.
We have approximated the power spectrum as a piece-wise constant and fit
the values of that constant within each bin to the combined data, taking
into account non-symmetric error bars and calibration uncertainties in
a manner similar to (12).
We maximize the likelihood function for the 8 parameters (plus 17 calibrations)
using a simulated annealing technique (13).
{}From the maximum likelihood position we then use Monte-Carlo integration to
calculate the covariance matrix of the parameters.
The final result is a power spectrum, with realistic estimates of the error
bars and bin-to-bin correlations.  We show the points and errors in
Figure~\ref{fig:bandpower}, and present the values in Table~\ref{tab:bins}.

\begin{table}[h] 
\centering
\begin{tabular}{cccc}
   $\ell_{\rm min}$ & $\ell_{\rm max}$ & $\ell(\ell+1)C_\ell/2\pi$ &
 $\pm1\sigma$ \\
   &       &        ($\mu\,{\rm K}^2$) & ($\mu\,{\rm K}^2$) \\
\hline
  2&  7 & 639 & 152 \\
  8& 15 & 814 & 160 \\
 16& 49 & 1048 & 298 \\
 50& 99 & 1394 & 367 \\
100& 149 & 3084 & 597 \\
150& 249 & 6548 & 590 \\
250& 449 & 2678 & 551 \\
450& 999 & 1971 & 825 \\
\end{tabular}
\caption{Band-powers and error bars plotted in
Figure~\protect\ref{fig:bandpower}.}
\label{tab:bins}
\end{table}

These points are somewhat correlated, with the strongest correlation being
typically a 30\% anti-correlation with immediately neighbouring bins, and
more distant correlations being almost negligible.
Table~\ref{tab:corr} explicitly shows the correlations between the
difference bins, fixing the calibrations at the maximum likelihood value.
Any use of these binned poewr spectrum estimates to constrain cosmological
models should include these correlations.
Our best fitting model has
$-2\ln{\cal L}=78$, a marginally acceptable fit.  We note that if the
experimental calibrations were not allowed to float, then the overall
$\chi^2$ would be far from acceptable.  In fact we find that the best fitting
calibration scalings are very close to unity for most experiments, with
the most discrepant values being 0.76 for MAT97, 0.83 for QMAT, 1.15 for MSAM
and 1.11 for BOOM97.

\begin{table}[h] 
\centering
\begin{tabular}{cccccccccc}
Bins  && 2--7 & 8--15 & 16--49 & 50--99 & 100-- & 150-- &
     250-- & 450-- \\
   &&       &        &         &         & 149  & 249  &
     449 & 999 \\
\\
  2--7 &&  1.00 & ---   & ---   & ---   & ---  &  --- &  --- & --- \\
  8--15 && -0.02 & 1.00 & ---   & ---   & ---  &  --- &  --- & --- \\
 16--49 && -0.04 &-0.08 & 1.00 & ---   & ---  &  --- &  --- & --- \\
 50--99 &&  0.02 & 0.03 &-0.33 & 1.00 & ---  &  --- &  --- & --- \\
100--149 &&  0.01 &-0.01 & 0.05 &-0.42 & 1.00&  --- &  --- & --- \\
150--249 && -0.00 & 0.00 &-0.04 & 0.07 &-0.41& 1.00&  --- & --- \\
250--449 &&  0.01 &-0.01 & 0.01 &-0.01 & 0.04&-0.22& 1.00& --- \\
450--999 &&  0.06 & 0.08 & 0.01 & 0.02 & 0.01& 0.05&-0.26&1.00\\
\end{tabular}
\caption{Correlations between the 8 bins shown in
Figure~\protect\ref{fig:bandpower}.}
\label{tab:corr}
\end{table}

These data show a prominent, localized peak in the angular power spectrum.
There is a distinct fall-off at high $\ell$, which is indicated within the
data sets of individual experiments (particularly Saskatoon (14), MAT, Viper
and BOOM97), but is more dramatically revealed in this compilation of data
sensitive to different angular scales.
Further confidence in the decrease in power comes from upper limits at even
larger $\ell$, not plotted or used in our fit.

In other words, there is a particular angular scale on which CMB temperature
fluctuations are highly correlated and that scale is around $\ell=200$,
or $0\degrees5$.
It corresponds theoretically to the distance a sound wave can have traveled
in the age of the Universe when the CMB anisotropies formed.
Such a characteristic scale was suggested in models of cosmological structure
formation at least as far back as 1970 (15).

The field is now in an exciting phase, with two main parts:
(a) confirming/refuting the basic paradigm; and (b) constraining the
parameters within that paradigm.  These go hand in hand, of course.
The peak prominent in Figure~1 confirms our ideas of the early evolution of
structure.  Understanding the physical basis for the peak allows a constraint
to be placed on the curvature of the universe (e.g.~16, 17).
The overall geometry of space appears to be close to flat, indicating that
something other than normal matter contributes to the energy density of the
Universe.  Together with data from distant supernovae and other cosmological
tests, this implies that models with cold dark matter and Einstein's
cosmological constant are in good shape (18).

Soon the detailed structure of the CMB spectrum should be measurable and we
expect it will contain a series of peaks and troughs.
Finding such structure in the spectrum at the correct $\ell$s would be
strong confirmation for `adiabatic' fluctuations (which perturb matter and
radiation in a similar way) produced at very early times.
Eventually this would lead to the possibility of `proving' inflation, or
stimulating research on other ways of generating similar fluctuations on
apparently acausal scales.
Of course, failure to see multiple peaks in the predicted locations would
require theorists to be more imaginative!

If we verify the framework we then need to determine precisely the parameters
within our model; namely the amounts of matter of different types, the
expansion rate, the precise form of the initial conditions, etc.
With a well characterized set of initial conditions we will clearly wish to
extend our understanding of cosmic origins to more recent epochs.
Even here the upcoming high resolution maps of the CMB will play a crucial
role carrying imprints, through reionization and gravitational lensing, of
object formation in the recent universe.

The future remains bright.  New results from a long duration flight of the
BOOMERANG experiment are expected in the very near future.  There are also
several ground-based experiments, including interferometric instruments,
nearing completion.
NASA's Microwave Anisotropy Probe is expected to return data in 2001, and the
ambitious Planck satellite is scheduled for launch in 2007.
Beyond this, information from challenging CMB polarization measurements and
the combination of CMB data with other cosmological probes will be even more
powerful.

We are on the threshold of precision measurements of the global properties
of our Universe.
The history of CMB research can be split into 5 phases.
Firstly, its mere existence showed that the early Universe was hot and dense.
Secondly, the blackbody nature of the CMB spectrum and its isotropic
distribution imply that the Universe is approximately homogeneous on large
scales.
The third step came with the detection of anisotropies which confirmed the
theory of structure formation through gravitational instability.
Here we have outlined a fourth stage, which is the discovery of a
characteristic (angular) scale on the CMB sky.
This supports a model with adiabatic initial conditions and a Universe with
approximately flat geometry.
Higher fidelity data, of the sort which will soon be available, should decide
whether or not our models are vindicated.
And now we are on the verge of the fifth phase, which involves determining the
precise values of the fundamental cosmological parameters to figure out
exactly what kind of Universe we live in.

\vspace{0.1in}

\hrule

\vspace{0.5in}

\begin{small}


\noindent
(1) A. A. Penzias, R. W. Wilson, {\it Astrophys. J.} {\bf 142}, 1149 (1965).

\noindent
(2) G. F. Smoot, et al., {\it Astrophys. J.} {\bf 396}, L1 (1992).

\noindent
(3) D. Scott, J. Silk, M. White, {\it Science} {\bf 268}, 829 (1995).

\noindent
(4) Technically, one takes
$\Delta T(\theta,\phi)=\sum_{\ell,m}\, a_{\ell m}Y_{\ell m}(\theta,\phi)$,
and plots $\ell(\ell+1)C_\ell/2\pi$ vs $\ell$, where
$C_\ell\equiv\left\langle|a_{\ell m}|^2\right\rangle/(2\ell+1)$.

\noindent
(5) L. Knox, {\it Phys. Rev.} {\bf D60}, 103516 (1999) [astro-ph/9902046].

\noindent
(6) G. F. Smoot, D. Scott, in {\it Review of Particle Properties}, C. Casi,
et al., {\it Eur. Phys. J.} {\bf C3}, 127 (1998) [astro-ph/9711069].

\noindent
(7) QMAP:
A. de Oliveira-Costa, et al., {\it Astrophys. J.} {\bf 509}, L77 (1998)
[astro-ph/9808045].

\noindent
(8) MAT:
E. Torbet, et al., {\it Astrophys. J.} {\bf 521}, L79 (1999)
[astro-ph/9905100];
A.D. Miller, et al., {\it Astrophys. J.} {\bf 524}, L1 (1999)
[astro-ph/9906421].

\noindent
(9) Viper:
J.B. Peterson, et al.,  {\it Astrophys. J.}, submitted [astro-ph/9910503].

\noindent
(10) BOOM97:
P.D. Mauskopf, et al., {\it Astrophys. J.}, submitted [astro--ph/9911444].

\noindent
(11) We are grateful to Lloyd Knox for making available his {\sc Radpack\/}
package\newline
\noindent
({\tt http://flight.\-uchicago.\-edu/\-knox/\-radpack.html}),
which we adapted for our analysis.

\noindent
(12) J. R. Bond, A. H. Jaffe, L. E. Knox, {\it Astrophys. J.}, in press
[astro-ph/9808264].

\noindent
(13) S. Hannestad, {\it Phys. Rev.\/} {\bf D61}, 023002 (2000)
[astro-ph/9911330].

\noindent
(14) Saskatoon: C. B. Netterfield, M. J. Devlin, N. Jarosik, L. Page,
E. Wollack, {\it Astrophys. J.} {\bf 474}, 47 (1997) [astro-ph/9601197].

\noindent
(15) P. J. E. Peebles, J. T. Yu, {\it Astrophys. J.} {\bf 162}, 815 (1970).

\noindent
(16) S. Dodelson, L. Knox, {\it Phys. Rev. Lett.}, in press
[astro-ph/9909454].

\noindent
(17) A. Melchiorri, et al., {\it Astrophys. J.}, in press [astro-ph/9911445].

\noindent
(18) N. A. Bahcall, J. P. Ostriker, S. Perlmutter, P. J. Steinhardt,
{\it Science} {284}, 1481 (1999) [astro-ph/9906463].

\noindent
This research was supported by the Natural Sciences and Engineering Research
Council of Canada and by NSF-9802362.


\noindent Authors e-mail addresses: 
elena@\-astro.\-ubc.\-ca;
dscott@\-astro.\-ubc.\-ca;
mwhite@cfa.\-harvard.\-edu
\end{small}

\clearpage

\end{document}